\documentclass[prx, notitlepage, twocolumn, nofootinbib, superscriptaddress]{revtex4}
\usepackage[colorlinks=true,citecolor=blue,linkcolor=blue]{hyperref}
\usepackage{graphicx}
\usepackage{color}
\usepackage{dcolumn}
\usepackage{bm}
\usepackage{mathrsfs}
\usepackage{amsmath}
\usepackage{amssymb}
\usepackage{amsthm}
\usepackage{amsfonts}
\usepackage{enumerate}
\usepackage{latexsym}
\usepackage{hyperref}
\usepackage{centernot}
\usepackage{multirow}
\usepackage{booktabs}
\usepackage{float}
\usepackage{multibib}
\newcites{M}{Methods References}
\begin{document}

\title{Magnetic Structures Database from Symmetry-aided High-Throughput Calculations}
\date{\today }
\author{Hanjing Zhou}
\thanks{These authors contributed equally to this work.}
\affiliation{National Laboratory of Solid State Microstructures and School of Physics,
Nanjing University, Nanjing 210093, China}
\affiliation{Collaborative Innovation Center of Advanced Microstructures, Nanjing
University, Nanjing 210093, China}
\author{Yuxuan Mu}
\thanks{These authors contributed equally to this work.}
\affiliation{National Laboratory of Solid State Microstructures and School of Physics,
Nanjing University, Nanjing 210093, China}
\affiliation{Collaborative Innovation Center of Advanced Microstructures, Nanjing
University, Nanjing 210093, China}
\author{Dingwen Zhang}
\thanks{These authors contributed equally to this work.}
\affiliation{National Laboratory of Solid State Microstructures and School of Physics,
Nanjing University, Nanjing 210093, China}
\affiliation{Collaborative Innovation Center of Advanced Microstructures, Nanjing
University, Nanjing 210093, China}
\author{Yuhui Li}
\thanks{These authors contributed equally to this work.}
\affiliation{Department of Physics and Guangdong Basic Research Center of Excellence for Quantum Science, Southern University of Science and Technology, Shenzhen 518055, China.}
\author{Xiaobing Chen}
\thanks{These authors contributed equally to this work.}
\affiliation{Quantum Science Center of Guangdong-Hong Kong-Macao Greater Bay Area (Guangdong),  Shenzhen 518045, China. }
\author{Hangbing Chu}
\affiliation{National Laboratory of Solid State Microstructures and School of Physics,
Nanjing University, Nanjing 210093, China}
\affiliation{Collaborative Innovation Center of Advanced Microstructures, Nanjing
University, Nanjing 210093, China}
\author{Fan Jiang}
\affiliation{National Laboratory of Solid State Microstructures and School of Physics,
Nanjing University, Nanjing 210093, China}
\affiliation{Collaborative Innovation Center of Advanced Microstructures, Nanjing
University, Nanjing 210093, China}
\author{Erjun Kan}
\affiliation{MIIT Key Laboratory of Semiconductor Microstructure and Quantum Sensing, Nanjing University of Science and Technology, Nanjing, Jiangsu 210037, People's Republic of China}
\author{Chun-Gang Duan}
\affiliation{Key Laboratory of Polar Materials and Devices, Ministry of Education, East China Normal University, Shanghai 200241, China}
\affiliation{Shanghai Center of Brain-inspired Intelligent Materials and Devices, East China Normal University, Shanghai 200241, China}
\author{Di Wang}
\thanks{The corresponding author: diwang0214@nju.edu.cn.}
\affiliation{National Laboratory of Solid State Microstructures and School of Physics,
Nanjing University, Nanjing 210093, China}
\affiliation{Collaborative Innovation Center of Advanced Microstructures, Nanjing
University, Nanjing 210093, China}
\author{Huimei Liu}
\thanks{The corresponding author: huimeiliu@nju.edu.cn.}
\affiliation{National Laboratory of Solid State Microstructures and School of Physics,
Nanjing University, Nanjing 210093, China}
\affiliation{Collaborative Innovation Center of Advanced Microstructures, Nanjing
University, Nanjing 210093, China}
\author{Qihang Liu}
\thanks{The corresponding author:  liuqh@sustech.edu.cn.}
\affiliation{Department of Physics and Guangdong Basic Research Center of Excellence for Quantum Science, Southern University of Science and Technology, Shenzhen 518055, China.}
\affiliation{Quantum Science Center of Guangdong-Hong Kong-Macao Greater Bay Area (Guangdong),  Shenzhen 518045, China.}
\affiliation{Guangdong Provincial Key Laboratory of Computational Science and Material Design, Southern  University of Science and Technology, Shenzhen 518055, China.}
\author{Xin-Gao Gong}
\affiliation{Key Laboratory of Computational Physical Sciences (Ministry of Education), Institute of Computational Physical Sciences,
and Department of Physics, Fudan University, Shanghai 200433, China}
\affiliation{Hefei National Laboratory, Hefei 230088, China}
\author{Xiangang Wan}
\thanks{The corresponding author: xgwan@nju.edu.cn.}
\affiliation{National Laboratory of Solid State Microstructures and School of Physics,
Nanjing University, Nanjing 210093, China}
\affiliation{Collaborative Innovation Center of Advanced Microstructures, Nanjing
University, Nanjing 210093, China}
\affiliation{Hefei National Laboratory, Hefei 230088, China}
\affiliation{Jiangsu Physical Science Research Center, Nanjing University, Nanjing
210093, China}

\begin{abstract}

Magnetic structures---the spatial arrangement of magnetic moments---underpin the diverse functionalities of magnetic materials, yet they have been experimentally determined for only a small fraction of known compounds. Conventional first-principles prediction requires candidate magnetic configurations as input, but the number of possible configurations is in principle infinite. Although the classic Landau-theory based symmetry analysis is widely used to classify magnetic ordering, their direct application to generate input magnetic configurations remains a major challenge due to unavoidable residual degrees of freedom. Here we turn this classification language into a practical predictive framework by resolving these freedoms through a two-step symmetry-aided strategy: an irrep-compatible maximal-magnetic-subgroup selection, followed by the physically motivated high-symmetry-direction construction. This highly efficient scheme is justified within the newly developed oriented spin space group framework. Benchmarked against 1,043 Landau-compatible experimental structures in MAGNDATA, our candidate sets reproduce 90\% of the experimentally reported magnetic structures in the four high-symmetry crystal systems with only $\sim$51 candidates per material on average, and subsequent first-principles calculations with material-specific Hubbard $U$ parameters recover the experimental ground state with 83\% accuracy. Applying the validated workflow to 19,808 transition-metal compounds, we construct a database of more than 5,000 magnetic materials---nearly twice the size of the experimental MAGNDATA. The database reveals statistical regularities accessible only at scale, including the occurrence rate of magnetism, the prevalence of competing magnetic states, and enables the identification of hundreds of magnetic topological materials, altermagnets, and other emerging magnetic phases and functionalities.

\end{abstract}

\maketitle

The essential characteristic of a magnetic material is its magnetic structure, which determines the magnetic symmetry and thereby constrains all of its physical properties, from transport and optical responses to collective excitations \cite{introduction-CMR,introduction-permanent,introduction-magtopo,introduction-multiferroic,introduction-spintronics,introduction-magnon,Broholm-SCIENCE2020,introduction-frustration,introduction-transition,PhysRevX.12.040501,PhysRevX.12.031042,cullity2011introduction}. Establishing the magnetic structure is thus a prerequisite for understanding and designing magnetic phenomena. Experimentally, magnetic structures are commonly probed by neutron diffraction, but reliable refinement is often limited by sample quality, crystal size, absorption, and the intrinsic ambiguity of powder data in fixing moment orientations \cite{Neutron-book,yankova2012crystals}. As a result, while the Inorganic Crystal Structure Database (ICSD) \cite{ICSD} records tens of thousands of compounds hosting magnetic ions, the experimental magnetic-structure database, i.e., MAGNDATA \cite{gallego2016magndata,gallego2016magndata-2}, covers only about two thousands of them, making theoretical prediction indispensable.

Theoretical prediction of magnetic structures generally follows a two-step paradigm, in which a set of initial magnetic configurations is first generated and first-principles total-energy calculations are then performed to identify the lowest-energy configuration. While first-principles total-energy methods can be continuously improved to rank candidates, such ranking is powerless if the experimentally realized configuration is absent from the candidate set. The central difficulty thus lies in generating the candidate set, because the number of conceivable magnetic structures for any given compound is in principle infinite. Existing generation strategies include constrained searches over collinear configurations \cite{horton2019high}, genetic-algorithm optimization \cite{zheng2021maggene}, iterative optimization guided by linear spin-wave theory \cite{tellez2024systematic}, spin-spiral techniques capturing long-wavelength noncollinear order \cite{sodequist2024magnetic}, and the cluster-multipole framework benchmarked on several hundred compounds \cite{huebsch2021benchmark,AritaPRB,Xu_MATEDATA}. These approaches have advanced the field within their respective scopes, yet the goal of database-scale magnetic-structure prediction continues to pose a challenge. The overwhelming majority of the potential magnets in ICSD thus remain unexplored even theoretically, calling for a practical candidate-generation strategy that combines high coverage of experimentally observed structures with candidate sets compact enough for high-throughput calculations.

Symmetry offers a complementary starting point. Most magnetic phase transitions are continuous and fall within the Landau framework, where the ordered phase is connected to the parent crystal through a group-subgroup relation and the instability transforms as an irreducible representation (irrep) of the parent space group \cite{authier2014international,SM}. Representation analysis and magnetic space groups (MSGs) provide a standard classification language\cite{magneticrepresentation1968,Neutron-book}, routinely used in neutron diffraction studies and implemented in tools such as ISODISTORT and MAXMAGN \cite{ISODISTORT,perez2015symmetry,RODRIGUEZCARVAJAL199355,Petricek2014}. Yet classification is not prediction: even with the relevant irrep specified, residual freedoms remain in the order-parameter direction for multidimensional irreps and in basis-vector combinations for repeated irreps, so that a bare irrep analysis uniquely determines the magnetic structure for only a small fraction of materials, as shown later.

Here we resolve these residual freedoms through a physically intuitive, easily implementable, and group-theoretically justified, two-level scheme. At the first level, irrep-compatible maximal magnetic subgroups fix the order-parameter direction of multidimensional irreps; this alone resolves $\sim$30\% of the benchmark cases. At the second level, moment orientations are sampled along the high-symmetry crystallographic directions of the parent structure compatible with the irreps, thereby reducing the remaining basis-vector-combination freedom associated with repeated irreps to a finite set of candidate magnetic structures. In practice, this scheme reduces the otherwise continuous search space to only tens of representative configurations per material while preserving high coverage ($\sim$75\%) of experimentally observed magnetic structures, making it suitable for database-scale calculations. The recently developed oriented spin space group (OSSG) framework \cite{liu2026symmetry} provides a systematic group-theoretical interpretation of this construction by incorporating additional spin-space symmetries that further reduce the dimensionality of the unresolved irrep subspace. Encouraged by the successful benchmark against more than 1,000 experimental magnetic structures in MAGNDATA, we construct a theoretical database of more than 5,000 predicted magnetic materials ($\sim$200,000 candidate magnetic structures), freely accessible at https://magdatabase.nju.edu.cn. By providing explicit magnetic structures, our database serves as a foundation for systematic studies of various facets of unconventional magnetism, including spintronics, magnonics, magnetic topology, and multiferroicity.

\section*{Theory and workflow}

\begin{figure*}[!tbp]
\centering

\includegraphics[width=0.98\textwidth]{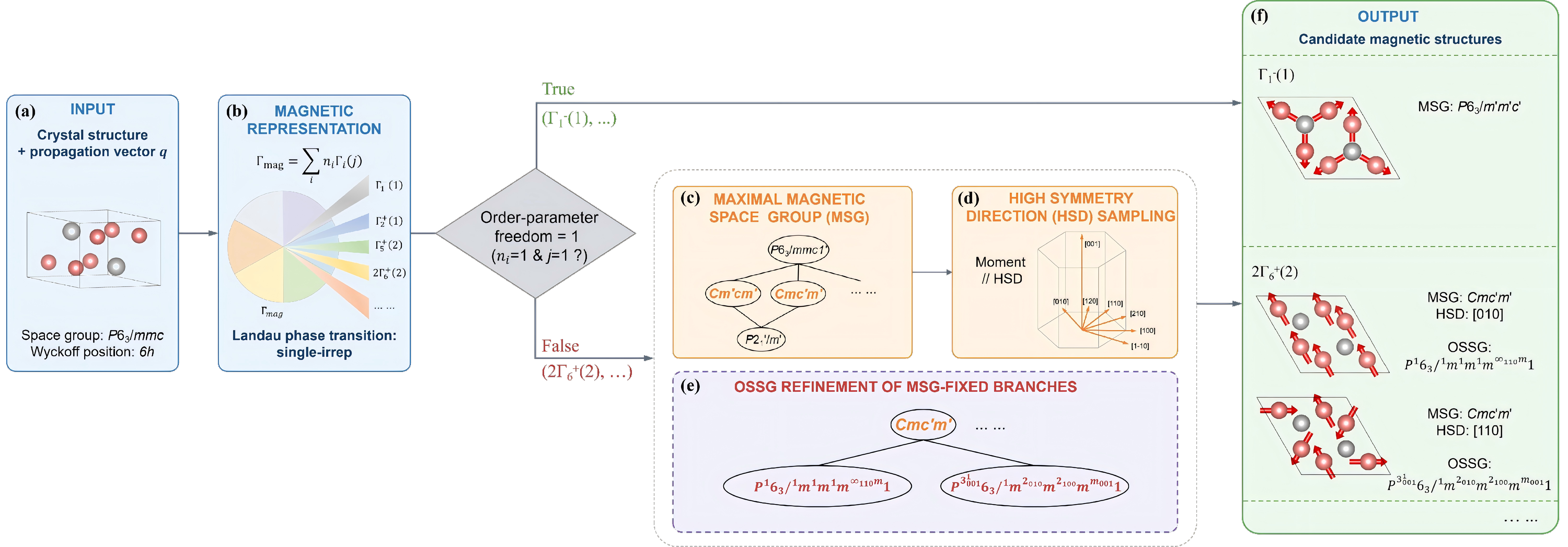}

\caption{\textbf{Symmetry-guided workflow for generating candidate magnetic structures.} Starting from a parent crystal structure, Wyckoff positions of magnetic ions and a target propagation vector \textbf{q}, the magnetic representation \(\Gamma_{mag}\) is constructed and decomposed into irreps of the little group of \textbf{q}. Under the Landau phase transition, when symmetry constrains the order-parameter freedom to 1, an explicit magnetic structure is directly locked, as shown in the upper branch. For cases with residual freedoms, the lower branch resolves continuous parameters by integrating maximal MSGs and the high-symmetry-direction approach, which can be supported by the maximal OSSGs that provide a more precise symmetry classification of magnetic structures. The generation of candidate magnetic structures is illustrated using Mn\textsubscript{3}Sn as a representative example, \textbf{a}, crystal structure of Mn\textsubscript{3}Sn with space group $P6_3/mmc$ (No. 194). \textbf{b}, irrep decomposition of the magnetic representation of Mn\textsubscript{3}Sn. \textbf{c}, hierarchy of irrep-compatible MSGs for the $\Gamma_{6}^{+}$ irrep. \textbf{d}, high-symmetry crystallographic directions of $P6_3/mmc$ used to restrict the magnetic-moment orientation when $n_i>1$. \textbf{e}, hierarchy of OSSGs associated with the  $\Gamma_{6}^{+}$ irrep. \textbf{f}, representative candidate magnetic structures obtained from the symmetry-guided construction strategy.}
\label{workflow}
\end{figure*}

The symmetry-guided workflow is rooted in Landau theory, in which a continuous magnetic phase transition is driven by a single active irrep (Fig. \ref{workflow}). Starting from a parent crystal structure, Wyckoff positions of magnetic atoms, and a target propagation vector \textbf{q}, we construct the magnetic representation \(\Gamma_{mag}\) and decompose it into irreps of the little group of \textbf{q},
\begin{equation}
\Gamma_{mag} = \sum_{i}^{}{n_{i}\Gamma_{i}(j),}\tag{1}
\label{eq:1}
\end{equation}
where \(\Gamma_{i}(j)\) denotes the \emph{i}-th irrep of dimension \emph{j}, and \(n_{i}\) is its multiplicity. For each \(\Gamma_{i}(j)\), the magnetic structure is parameterized by an order-parameter vector \(\eta = \left( \eta_{1},\ ...,\eta_{j} \right)\) and symmetry-adapted basis vectors \(\varphi_{\gamma}\), i.e., \(M = \sum_{\gamma=1}^{j}\eta_{\gamma}\varphi_{\gamma}\). When \(n_{i} = j = 1\), symmetry uniquely determines the magnetic structure within the conventional Landau-irrep framework. However, in most materials, a multidimensional irrep (i.e., \(j > 1\)) or repeated irrep (i.e., \(n_{i} > 1\)) leaves residual freedom in the order-parameter direction or basis-vector combinations, respectively.

The proposed two-level scheme addresses this challenge by systematically constraining these continuous spaces. At the first level, for a multidimensional irrep, we enumerate the irrep-compatible maximal magnetic subgroups of the parent grey group \cite{perez2015symmetry,Zhou-FeGe,Ascher1977}. Each maximal MSG defines a fixed subspace in the order-parameter space and therefore restricts the otherwise continuous order-parameter direction. At the second level, for the case where the maximal MSG construction leaves residual freedom, typically arising from repeated irreps, we introduce a high-symmetry-direction approximation: we approximate the remaining continuous freedom by sampling high-symmetry crystallographic directions compatible with the irrep basis vectors, thereby reducing it to a finite set of candidate structures.

For example, Mn$_3$Sn crystallizes in space group \(P6_{3}/mmc\), with the magnetic Mn ions occupying the $6h$ Wyckoff position. Here we use \textbf{q} = 0 as an illustrative case, while the extension to \textbf{q} \(\neq\) 0 is straightforward (see Section I of SM \cite{SM}). Its magnetic representation contains ten irreps (see Section I of SM \cite{SM}). For the one-dimensional irreps that occur only once, such as \(\Gamma_{1}^{-}(1)\), the magnetic structure is uniquely determined, as illustrated in Fig. \ref{workflow}\textbf{f}. For a multidimensional irrep such as \(\Gamma_{6}^{+}(2)\), the maximal-MSG construction yields the two compatible maximal magnetic subgroups $Cm^\prime cm^\prime$ and $Cmc^\prime m^\prime$, as shown in Fig. \ref{workflow}\textbf{c}. Because \(\Gamma_{6}^{+}(2)\) occurs twice, however, even after selecting $Cmc^\prime m^\prime$, a two-dimensional unresolved basis-vector subspace remains. We further constrain the moment orientation along $\left\langle 010\right\rangle$ and $\left\langle 110\right\rangle$ directions (among seven typical high-symmetry directions of the hexagonal crystal system) that are compatible with the basis vectors of the \(\Gamma_{6}^{+}(2)\) irrep, thereby determining two magnetic configurations: the collinear ferromagnetic (FM) state and the 120° coplanar antiferromagnetic (AFM) state, as shown in Fig. \ref{workflow}\textbf{f}. The latter is precisely the experimentally reported magnetic structure of Mn$_3$Sn\cite{Brown1990,cederholm2026groundstatemagneticstructure}, confirming that our two-level scheme correctly captures the ground state.

The success of the two-level scheme is systematically justified in the OSSG framework, which provides a comprehensive description of both the magnetic geometry and symmetry breaking driven by spin–orbit coupling (SOC). In the conventional MSG description, spin rotations are locked to crystallographic operations, so that physically distinct magnetic configurations can share the same MSG label. The OSSG framework overcomes this limitation by treating spin and spatial rotations as independent operations and explicitly fixing the spin orientation relative to the crystal lattice. This distinguishes configurations that share identical MSG label. The additional spin-space symmetries in the OSSG description can further reduce the dimensionality of the unresolved order-parameter spaces in Eq. (\ref{eq:1}), thereby refining the candidate set beyond what the maximal-MSG construction alone can achieve.

For example, in the \(2\Gamma_{6}^{+}(2)\) irrep subspace discussed above, the maximal-MSG construction leaves a two-dimensional unresolved order-parameter space, rendering states sharing the same MSG $Cmc^\prime m^\prime$ but different magnetic configurations indistinguishable. The OSSG framework resolves this continuous ambiguity by decomposing the two-dimensional subspace into discrete one-dimensional symmetry-adapted branches, unambiguously distinguishing the collinear ferromagnet from the 120° coplanar antiferromagnet. When structural complexity leaves residual continuous freedom even under OSSG resolution, sampling high-symmetry crystallographic directions physically extends the workflow's predictive coverage to complex magnets that group theory alone leaves undetermined. Details of the magnetic-representation construction, irrep-compatible maximal-MSG selection, high-symmetry-direction approximation, and OSSG refinement are provided in Methods.

\section*{Benchmark and validation}

We validate the symmetry-guided workflow through a two-stage benchmark against experimentally reported magnetic structures in the MAGNDATA database \cite{gallego2016magndata,gallego2016magndata-2}. After removing 628 duplicate entries, 346 fractional site occupancies, 97 incommensurate orders and 70 reported multi-\textbf{q} structures, 1,187 commensurate entries remain. Among these, 1,083 (91\%) are compatible with a single-irrep Landau theory, and 1,043 of the 1,083 entries (96\%) possess propagation vectors at high-symmetry \textbf{q} points, where all three components are fixed by symmetry with no free parameters (see Section II of Supplementary Information \cite{SM}), forming the benchmark set. The remaining entries involving multi-irrep couplings or generic \textbf{q} points lie beyond the present single-irrep scope and represent natural extensions of the framework.

The first benchmark evaluates whether the generated candidate sets contain the experimentally observed magnetic structure, assessed within a 5° angular tolerance of the moment direction. A bare Landau-irrep analysis can generate a complete finite candidate set only when all relevant irreps are one-dimensional and occur with unit multiplicity, a rather restrictive condition satisfied by only 62 of the 1,043 benchmark materials ($\sim$6\%). Incorporating the irrep-compatible maximal-MSG further constrains the order-parameter freedom of multidimensional irreps and fully resolves the magnetic structures of 313 materials (30.0\%). Remarkably, sampling moment orientations along high-symmetry crystallographic directions brings the total to 782 structures, achieving 75\% coverage within the 5° tolerance (80\% at a 10° tolerance). The coverage exhibits a pronounced dependence on crystal symmetry, as summarized in Table \ref{tab:coverage}. For the four high-symmetry systems (tetragonal, trigonal, hexagonal, and cubic), an average of only 51 candidates per material achieves 90\% coverage (490 of 543). In contrast, the lower-symmetry triclinic, monoclinic, and orthorhombic systems yield substantially lower coverage, reflecting the reduced constraining power of crystallographic symmetry on moment orientations. These statistics provide a physically transparent reliability map: predictions for high-symmetry compounds carry intrinsically higher confidence.

\begin{table*}[htbp]
\caption{
Coverage of the candidate magnetic structure set across different crystal systems. The coverage is evaluated at different angular tolerances and is defined as the fraction of materials in the MAGNDATA benchmark for which the experimentally observed magnetic structure is reproduced by at least one candidate structure.The mean number of candidate magnetic structures per material is also listed for each crystal system.The crystal systems are grouped into high-symmetry systems, namely tetragonal, trigonal, hexagonal, and cubic, and low-symmetry systems, namely triclinic, monoclinic, and orthorhombic.
}
\label{tab:coverage}
\begin{ruledtabular}
\begin{tabular}{lccccc}
Crystal system & Number & Coverage & Coverage & Coverage & average \\
 & of materials & $(0^\circ)$ & $(5^\circ)$ & $(10^\circ)$ & per material \\
 Grand Total & 1043 & 68\% (707/1043) & \textbf{75\%} (782/1043) & 80\% (830/1043) & 57 \\
\hline
Triclinic & 10 & 20\% (2/10) & 20\% (2/10) & 20\% (2/10) & 24 \\
Monoclinic & 172 & 23\% (39/172) & 56\% (96/172) & 69\% (119/172) & 55 \\
Orthorhombic & 318 & 58\% (186/318) & 61\% (194/318) & 65\% (207/318) & 86 \\
Total & 500 & 45\% (227/500) & 58\% (292/500) & 66\% (328/500) & 70 \\
\hline
Tetragonal & 256 & 92\% (235/256) & 94\% (241/256) & 95\% (243/256) & 53 \\
Trigonal & 118 & 81\% (95/118) & 81\% (96/118) & 84\% (99/118) & 41 \\
Hexagonal & 66 & 95\% (63/66) & 95\% (63/66) & 95\% (63/66) & 56 \\
Cubic & 103 & 84\% (87/103) & 87\% (90/103) & 94\% (97/103) & 54 \\
Total & 543 & 88\% (480/543) & \textbf{90\%} (490/543) & 92\% (502/543) & \textbf{51} \\
\end{tabular}
\end{ruledtabular}
\end{table*}

The second benchmark evaluates whether DFT total-energy calculations correctly identify the experimental structure among the generated candidates. From the 782 materials whose experimental structures were successfully generated, we exclude 318 containing f-electron elements \cite{kotliar2006electronic} and 53 with more than 60 atoms per primitive cell, leaving 411 materials for the DFT benchmark. Candidate structures are evaluated using the Vienna \textit{Ab initio} Simulation Package (VASP)~\cite{vasp1,vasp2} within spin-polarized GGA+SOC+$U$ calculations at three Hubbard values ($U$ = 0, $U_{lr}$/2, $U_{lr}$) \cite{VASPUtypeDudarev1998,GGAU}, where $U_{lr}$ is obtained from linear response method\cite{PhysRevB.71.035105}.
A prediction is counted as correct if the experimental structure lies within 30 \ensuremath{\mu}eV per magnetic atom of the lowest-energy state for at least one $U$ value. Under this criterion, DFT successfully recovers the experimental ground state in 340 compounds, achieving 83\% accuracy. This accuracy is notably higher than that obtained with a single fixed $U$ value (see Methods and Supplementary Information \cite{SM}), underscoring the importance of material-specific correlation parameters.

Taken together, the two stages yield an overall prediction reliability of approximately 0.90 × 0.83 \ensuremath{\approx} 75\% for high-symmetry crystal systems---the regime where the majority of magnetic materials of current interest reside. These benchmarks, performed on around 1,000 experimental structures rather than a small set of selected examples, establish the quantitative foundation for database-scale magnetic-structure prediction.

\section*{High-throughput Magnetic Structure Database}

Encouraged by the high benchmark performance, we now apply the workflow to large-scale prediction. From the Inorganic Crystal Structure Database (ICSD) \cite{ICSD}, filtering out lanthanoid and actinoid compounds, fractional site occupancies, structures with 60 or more atoms per primitive cell, and duplicate entries leaves 19,808 stoichiometric transition-metal compounds. For each parent structure, all high-symmetry \textbf{q} points are evaluated, generating approximately {50} candidate magnetic structures per compound on average---nearly one million first-principles inputs in total. All candidates are calculated using spin-polarized GGA+SOC+$U$ with three Hubbard $U$ values ($U$ = 0, $U_{lr}$/2, $U_{lr}$), and the lowest-energy structure is assigned as the predicted ground state, with configurations within 30 \ensuremath{\mu}eV per magnetic atom retained as near-degenerate states. This procedure yields predicted ground-state magnetic structures for {5055} compounds, which are hosted at https://magdatabase.nju.edu.cn, along with MSGs, propagation vectors, magCIF files, material-specific Hubbard parameters, energy hierarchies, band structures, and computational inputs. Importantly, more than 4,600 of these compounds have no experimentally reported magnetic structures in MAGNDATA, substantially extending the magnetic-structure landscape beyond the existing experimental record. To our knowledge, this constitutes the largest magnetic-structure database to date: comparable in scope to the experimental MAGNDATA database while nearly doubling its size, it provides the first theory-derived counterpart against which essentially every newly measured magnetic structure can be cross-referenced. Notably, the predicted magnetic structures of CaMnSb$_2$ \cite{Li_2024}, Cs$_2$RuO$_4$ \cite{n8rr-rgfv}, Ba$_2$FeSbS$_5$ \cite{10.1021/acs.inorgchem.4c03770} and Ba$_2$FeBiS$_5$ \cite{10.1021/acs.inorgchem.4c03770}, which are not included in MAGNDATA, have been independently determined by neutron-diffraction experiments, with the measured magnetic orders showing good agreement with our predictions. These cases provide an independent validation of the predictive capability of our framework beyond the MAGNDATA benchmark.

\begin{figure*}[htbp]
\centering
\includegraphics[width=0.98\textwidth]{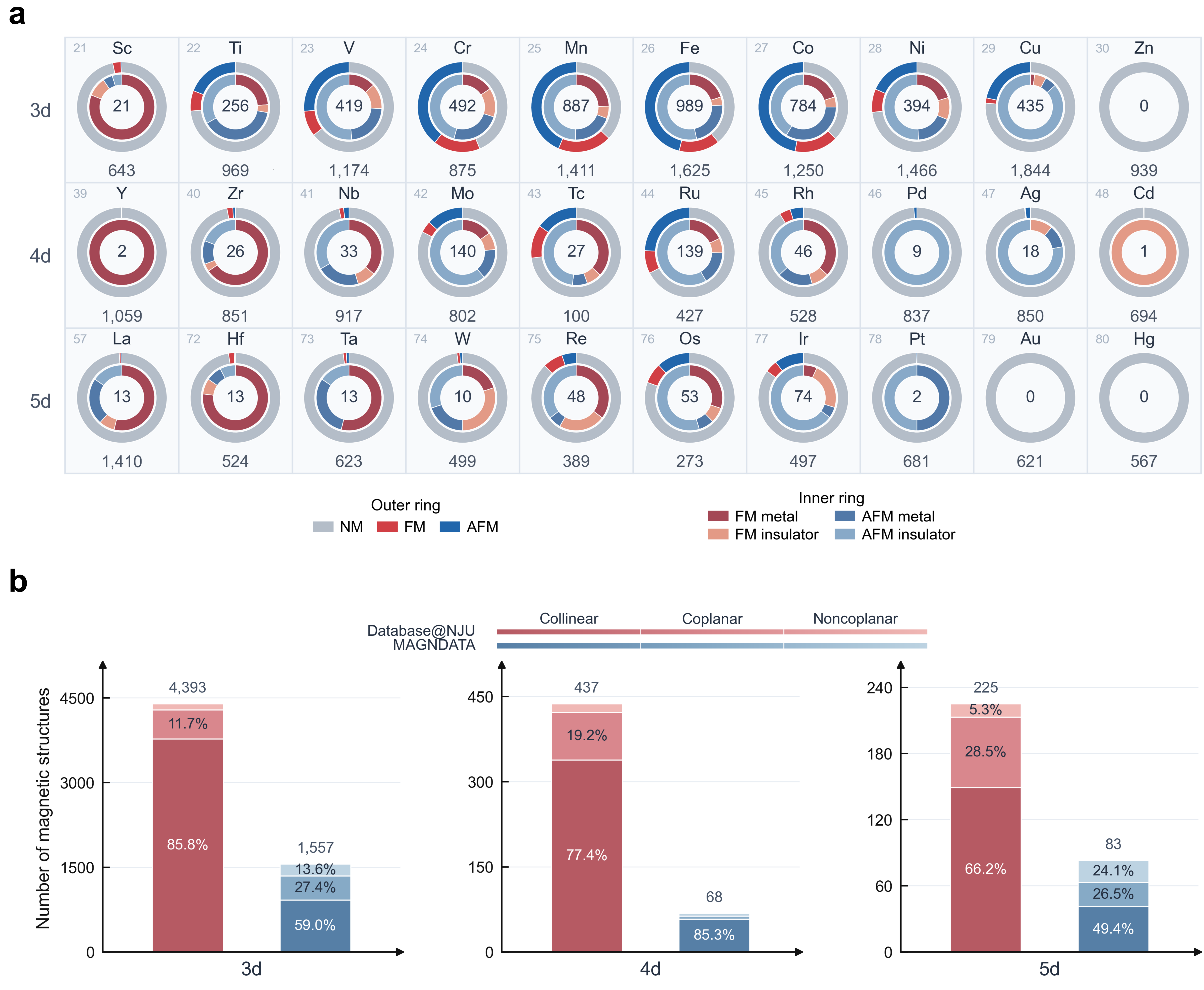}
\caption{ \textbf{Distribution of predicted magnetic ground states}. \textbf{a}, \textbf{Element-resolved distribution across the 3\emph{d}, 4\emph{d} and 5\emph{d} transition-metal series}. Numbers at the centers denote the occurrence counts of predicted magnetic ground states containing each element, while numbers below the rings specify the calculated crystal structures. Outer rings report the fractions of nonmagnetic (NM), FM, and AFM states. Inner rings further resolve the electronic structures within FM and AFM classes, distinguishing FM metals, FM insulators, AFM metals, and AFM insulators. \textbf{b}, \textbf{Comparison of collinear, coplanar and noncoplanar magnetic structures in predicted dataset with the experimental database MAGNDATA}. Relative fractions of collinear, coplanar, and noncoplanar magnetic structures across the 3\emph{d}, 4\emph{d}, and 5\emph{d} series, comparing high-throughput predictions against the experimental MAGNDATA database.}
\label{Distribution}
\end{figure*}

Fig. \ref{Distribution}\textbf{a} summarizes the distribution of {5055} predicted magnetic ground states evaluated at $U=U_{lr}$ across the 3\emph{d}, 4\emph{d}, and 5\emph{d} series, respectively. Across each transition metal series, the magnetic occurrence rate follows a characteristic volcano-shaped trend that peaks near mid-series elements. This underscores that the formation of magnetism in diverse crystals remains fundamentally governed by Hund\textquotesingle s first rule, in which maximizing the total spin quantum number in near-half-filled \emph{d} shells stabilizes local magnetic moments. For example, in the 3\emph{d} series, Fe, Mn and Co yield {989, 887, and 784} magnetic ground states out of 1,625, 1,411, and 1,250 crystal structures, respectively. In contrast, the magnetic occurrence rate drops sharply to below 5\% for most 4\emph{d} and 5\emph{d} elements because more delocalized 4\emph{d} and 5\emph{d} electrons produce larger bandwidths that fail the Stoner criterion in unit cells\cite{Stoner1938}. To establish magnetic order, a prominent majority of surviving 4\emph{d} and 5\emph{d} magnetic structures, representing $\sim$60\% of cases, adopt supercell AFM ordering with \textbf{q} \ensuremath{\neq} 0 to open a Slater pseudo-gap via band folding\cite{Slater1951}, which suppresses the effective bandwidth and stabilizes localized magnetic moments.

Among these predicted magnetic structures, AFM ordering predominates with {3,792} compounds, constituting {75\%} of the dataset, close to 76.6\% in the experimental database MAGNDATA. The remaining {1,263 FM-type systems comprise 620 ferromagnets, 518 ferrimagnets, and 125 compensated ferrimagnets}. Analysis of propagation vectors further demonstrates quantitative alignment with experimental observations, structures with non-zero propagation vectors account for {63.6\% (3,214/5,055)} of the dataset, matching the 61.3\% frequency of \textbf{q} \ensuremath{\neq} 0 structures in MAGNDATA. This dual quantitative agreement demonstrates that both the magnetic order ratio and propagation tendencies remain robust even an expanded material search space. Electronic structure calculations further illustrate the microscopic exchange mechanisms. Among the {3,064 insulating magnetic materials, 2,692 are predicted to be AFM, whereas only 372 are FM}, including {83 compensated ferrimagnets}. By contrast, the metallic systems comprise {1,100} antiferromagnets and {891} ferromagnets. In addition, we find that 33.6\% (1,696/5,055) of predicted magnetic structures have near-degenerate states within a narrow 30 \emph{\ensuremath{\mu}}eV per magnetic atom window, yielding highly tunable magnetic states via external strain or fields.

Fig. \ref{Distribution}\textbf{b} shows the distribution of collinear, coplanar and non-coplanar magnetic structures across the 3\emph{d}, 4\emph{d}, 5\emph{d} series. Collinear magnetic structures form the vast majority of predicted ground states, with {3,770, 338 and 149} occurrences in the 3\emph{d}, 4\emph{d} and 5\emph{d} series, respectively. On the other hand, the noncollinear fraction exhibits a clear upward trend from {14.2\% in 3\emph{d} to 33.8\% in 5\emph{d}.} This trend captures the progressively enhancing SOC in heavier elements. This SOC enhancement amplifies anisotropic exchange interactions, such as the Dzyaloshinskii-Moriya interaction\cite{DZYALOSHINSKY1958241,Moriya1960}, thereby supporting noncollinear magnetic structures. Systematic evaluations under different \emph{U} values are provided in the Supplementary Information\cite{SM}. This database addresses the long-standing scarcity of potential magnetic materials in spintronic applications, providing a fertile platform towards next-generation spintronic and quantum devices.

\section*{Properties of Predicted Magnets}

\begin{figure*}[htbp]
\centering
\includegraphics[width=0.98\textwidth]{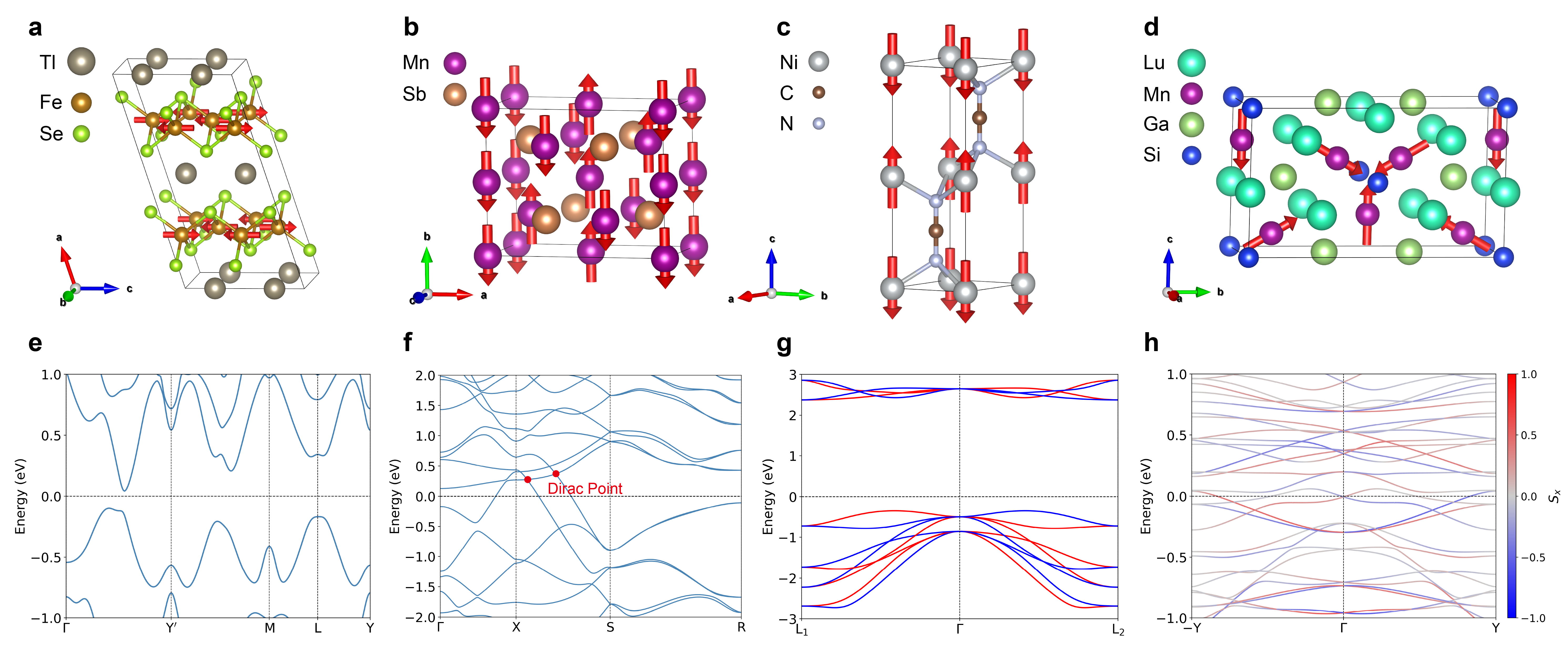}
\caption{ Magnetic structures and electronic band structures of four representative unconventional magnets. \textbf{a}--\textbf{d}, Magnetic structures of the AFM topological insulator TlFe\textsubscript{2}Se\textsubscript{2} (\textbf{a}), AFM Dirac semimetal Mn\textsubscript{2}Sb (\textbf{b}), altermagnet NiCN\textsubscript{2} (\textbf{c}) and \emph{p}-wave magnet Lu\textsubscript{3}Mn\textsubscript{3}Ga\textsubscript{2}Si (\textbf{d}). Red arrows indicate the directions of the magnetic moments. \textbf{e}--\textbf{h}, Corresponding electronic band structures in the same order. The red dot in \textbf{f} marks the Dirac point. In \textbf{g}, red and blue denote spin-up and spin-down bands, respectively; in \textbf{h}, the colour scale indicates the spin projection \emph{S\textsubscript{x}}, from negative (blue) to positive (red).}
\label{materials}
\end{figure*}

The proposed magnetic structure database enables property-oriented discovery at a scale that was previously impractical, addressing the long-standing scarcity of potential magnetic materials for next-generation spintronic and topological-based devices. From this database, we identified 927 topologically nontrivial magnetic materials, including 205 topological insulators and 722 topological semimetals, alongside 547 spin-split antiferromagnets and 325 spin-orbit magnets. The spin-split antiferromagnets include 338 altermagnets and 114 \emph{p}-wave magnets \cite{PhysRevX.12.031042,hellenes2024pwavemagnets}. A complete list of materials, together with their symmetry characteristics and physical properties, is provided in the Supplementary Information \cite{SM}.

As representative topological phases identified in the database, we present the AFM $\mathbb{Z}_2$ topological insulator TlFe\textsubscript{2}Se\textsubscript{2} (icsd\_166445), and the AFM Dirac semimetal Mn\textsubscript{2}Sb (icsd\_163332). TlFe\textsubscript{2}Se\textsubscript{2} crystallizes in the tetragonal space group \emph{I}4/\emph{mmm} (No. 139) with a collinear AFM ground state characterized by the propagation vector \textbf{q} = (1/2, 0, 1/2) and MSG $C_c2/c$ (No. 15.90) (Fig. \ref{materials}\textbf{a}). Its electronic structure exhibits the insulating behavior with a direct gap of 0.18 eV and an indirect bandgap of 0.13 eV (Fig. \ref{materials}\textbf{e}). The $T\tau_y$ symmetry protects a $\mathbb{Z}_2$ topological invariant, which is evaluated to be non-trivial ($\mathbb{Z}_2$ = 1) based on the parity eigenvalues at the \emph{T}-invariant momenta\cite{PhysRevB.81.245209}. Mn\textsubscript{2}Sb has a hexagonal parent structure with space group \emph{P}6\textsubscript{3}/\emph{mmc} (No. 194). Its predicted collinear AFM ground state is characterized by the propagation vector \textbf{q} = (1/2,0,0) and MSG \emph{P\textsubscript{B}nma} (No. 62.454) (Fig. \ref{materials}\textbf{b}). Its electronic band structure exhibits two Dirac points along the X-S path, protected by the $C_{2y}\tau_y$ symmetry (Fig. \ref{materials}\textbf{f}). The surface states and the robustness of both topological phases across various Hubbard $U$ parameter are presented in the Supplementary Information \cite{SM}.

Beyond topological phases, our database systematically identifies a wide spectrum of unconventional spin-split magnets, exemplified by the altermagnet NiCN\textsubscript{2} (icsd\_249388) and the \emph{p}-wave magnet Lu\textsubscript{3}Mn\textsubscript{3}Ga\textsubscript{2}Si (icsd\_099080). NiCN\textsubscript{2} crystallizes in the hexagonal space group \emph{P}6\textsubscript{3}/\emph{mmc} (No. 194) and exhibits \emph{A}-type AFM order with OSSG \(P^{\overline{1}}6_{3}/_{\ }^{\overline{1}}m_{\ }^{1}m_{\ }^{\overline{1}}c_{\ }^{\infty_{001}m}1\) (Fig. \ref{materials}\textbf{c}). In the absence of SOC, the bands exhibit large momentum-dependent spin splitting exceeding 0.4 eV near the Fermi level (Fig. \ref{materials}\textbf{g}). Lu\textsubscript{3}Mn\textsubscript{3}Ga\textsubscript{2}Si crystallizes in the hexagonal space group \(P\overline{6}2m\) (No. 189) with a coplanar AFM order in the \emph{bc} plane, which belongs to the OSSG \(A_{\ }^{1}m_{\ }^{2_{001}}m_{\ }^{2_{001}}2|\left( 1,1,1;2_{100} \right)_{\ }^{m_{100}}1\) (Fig. \ref{materials}\textbf{d)}. The presence of \emph{U}\textsubscript{2\emph{x}}\emph{\ensuremath{\tau}} forbids the $S_{y}$ and $S_{z}$ components, leaving $S_{x}$ as the only allowed spin component in the chosen spin frame. The $T\tau$ symmetry ($\tau$ is a \emph{bc}-plane fractional translation), relates opposite spin polarizations at \textbf{k} and -\textbf{k}, giving an odd-parity spin-splitting pattern. The calculated bands accordingly have opposite \emph{S\textsubscript{x}} projections on either side of $\Gamma$ along $-\mathrm{Y}$-$\Gamma$-$\mathrm{Y}$ (Fig. \ref{materials}\textbf{h}). Remarkably, this coplanar magnetic configuration lowers the non-polar crystal symmetry down to a polar one, making Lu\textsubscript{3}Mn\textsubscript{3}Ga\textsubscript{2}Si a candidate for electric manipulation of the spin polarization. Both magnetic structures and their spin-splitting patterns persist across the Hubbard \emph{U} values considered (Supplementary Information \cite{SM}).

\section*{Outlook}

In this work, we develop a symmetry-aided two-level framework that bridges the long-standing gap between Landau-irrep symmetry classification and practical magnetic-structure prediction. Based on this, we construct a large theoretical magnetic-structure database. Several natural extensions can further broaden the scope of the present framework. It can be generalized to multi-\(q\) magnetic orders and magnetic phases involving coupled irreps. For low-symmetry systems, including triclinic, monoclinic, and orthorhombic crystals, introducing additional moment directions in the second step of our strategy may further improve candidate-set coverage. More adaptive treatments of Hubbard \(U\) and electronic correlations may also enhance the reliability of magnetic-ground-state prediction.

Beyond these methodological extensions, a large-scale theoretical magnetic-structure database can bridge experimental structure determination and subsequent theoretical studies. For neutron-diffraction experiments\cite{magneticrepresentation1968,RODRIGUEZCARVAJAL199355}, particularly when powder data do not uniquely determine moment orientations, suitable single crystals are unavailable, or several magnetic configurations are nearly degenerate, the predicted candidate structures, propagation vectors, magnetic space groups, and energy hierarchy can substantially narrow the refinement space and provide useful theoretical guidance. For angle-resolved photoemission and scanning-tunneling experiments\cite{krempasky2024altermagnetic,PhysRevLett.132.036702}, comparison between calculated electronic structures and measured spectra may provide complementary information for identifying the underlying magnetic order.

The predicted ground states and metastable magnetic configurations also provide direct inputs for large-scale studies of magnetic topology, unconventional magnetism, spin-dependent transport, nonlinear optical responses, magnonics, and other magnetic functionalities\cite{xu2020high,PhysRevX.12.040501,AritaPRB,Chen2025,introduction-magnon2,Nagaosa2010,Sun2019}. They further offer natural starting points for constructing microscopic spin models, enabling systematic studies of magnetic excitations, magnetic frustration, and finite-temperature transitions. In particular, the large number of near-degenerate states opens a materials-scale route to identify compounds where modest strain, field, or compositional tuning can switch the magnetic ground state\cite{Cherifi2014,Tokura1996}, making them natural targets for studying competing magnetic orders and magnetoelectric coupling. Ultimately, we envision the database evolving from a catalog of magnetic ground states into a broader platform connecting crystal structures, magnetic orders, microscopic interactions, collective excitations, and functional properties, enabling a statistical, predictive approach to magnetic materials discovery.

\section*{Acknowledgements}

This work was financially supported by the National Natural Science Foundation of China (Grants No. 12188101, No. 12574066, No. 12474233, No. 12525410 and No. 12334007), the National Key R\&D Program of China (Grant No. 2025YFA1411301, No. 2022YFA1403601), Innovation Program for Quantum Science and Technology (Grant No.2021ZD0301902), Quantum Science and Technology-National Science and Technology Major Project (No. 2024ZD0300101), Natural Science Foundation of Jiangsu Province (No. BK20233001, No. BK20243011), Fundamental and Interdisciplinary Disciplines Breakthrough Plan of the Ministry of Education of China (No. JYB2025XDXM411), Guangdong Provincial Quantum Science Strategic Initiative under Grant No. GDZX2601003, the Fundamental Research Funds for the Central Universities (No. KG202501). This work has been supported by the New Cornerstone Science Foundation.

\newpage
\section*{Methods}

\subsection*{Symmetry-based generation of candidate magnetic structures}

A magnetic structure can be expanded in terms of the symmetry-adapted basis vectors of the irreps of the parent crystallographic space group,
\[\mathbf{M} = \sum_{i}^{}{\sum_{\gamma = 1}^{j}\eta_{\gamma}^{i}}\mathbf{\phi}_{\gamma}^{i}.\]
where \(\mathbf{\phi}_{\gamma}^{i}\) is the \(\gamma\)-th basis vector of the \(i\)-th irrep \(\Gamma_{i}(j)\), and \(\eta_{\gamma}^{i}\) is the corresponding order-parameter component. Determining a magnetic structure therefore requires identifying the relevant irrep, its symmetry-adapted basis vectors and their relative weights. For a continuous magnetic phase transition, the ordered magnetic phase is related to the parent crystallographic phase through a group--subgroup relation. Directly searching the corresponding magnetic subgroups, however, leads to a typically vast configuration space. Within Landau theory, a continuous transition is driven by the instability of a single irrep, providing a natural way to substantially restrict this search space. For each material, we first construct the magnetic representation from the magnetic moments associated with the magnetic Wyckoff positions. 
Here we use \textbf{q} = 0 as an illustrative case, while the extension to \textbf{q} \(\neq\) 0 is straightforward \cite{SM}.
For a unit cell containing \(N\) magnetic ions, the magnetic representation spans the \(3N\)-dimensional space defined by the moment components \(\left( m_{1x},m_{1y},m_{1z},\ldots,m_{Nx},m_{Ny},m_{Nz} \right)\). The resulting reducible magnetic representation is decomposed as
\[\Gamma_{mag} = \sum_{i}^{}{n_{i}\Gamma_{i}(j)}\]
where \(n_{i}\) is the multiplicity of \(\Gamma_{i}(j)\). Only the irreps appearing in this decomposition are relevant magnetic-ordering channels for the given magnetic Wyckoff positions, thereby substantially reducing the search space of input magnetic structures. The completeness relation
\[\sum_{i}^{}n_{i} \cdot j = 3N\]
ensures that all magnetic degrees of freedom are included. For a given irrep \(\Gamma_{i}(j)\), the conventional Landau-irrep construction uniquely determines the magnetic configuration only in the special case \(j = 1\) and \(n_{i} = 1\). In general, two residual degrees of freedom remain. For multidimensional irreps with \(j > 1\), the relative order-parameter components \(\eta_{\gamma}^{i}\) are not uniquely determined. When \(n_{i} > 1\), repeated copies of the same irrep can be linearly combined, making the corresponding basis-vector combinations nonunique.

\subsection*{Irrep-compatible maximal magnetic subgroups}

For a multidimensional irrep (\(j > 1\)), Landau theory identifies the unstable irrep but does not generally determine the specific direction of the order parameter, which depends on higher-order terms in the magnetic free energy. For a continuous phase transition, the ordered phase is physically expected to emerge through minimal symmetry breaking. Group-theoretically, this corresponds to the smallest reduction of the parent symmetry allowed by the unstable irrep, namely an irrep-compatible maximal magnetic subgroup.

We therefore enumerate the maximal magnetic subgroups compatible with each selected irrep. Each maximal subgroup preserves the largest possible residual symmetry and fixes a symmetry-distinguished direction in the order-parameter space, thereby converting the continuous freedom of a multidimensional irrep into a finite set of candidate magnetic space groups. Details of the subgroup construction and benchmark are provided in the Supplementary Information.

\subsection*{High-symmetry-direction approximation.}

When $n_i>1$, the same irrep appears multiple times in $\Gamma_{mag}$. The corresponding basis-vector sets can then be linearly combined with continuous coefficients. Consequently, even after the irrep and a compatible maximal MSG are specified, the explicit magnetic configuration may remain undetermined because different basis-vector combinations can produce different local-moment orientations.

In principle, a stable magnetic configuration requires the magnetic free energy to be minimized with respect to these angular degrees of freedom. If the direction of the local moment on site \emph{n} is described by an angle \emph{\ensuremath{\theta}\textsubscript{n}}, the equilibrium condition can be written schematically as
\[\frac{\partial F}{{\partial\theta}_{n}}\  = \ 0.\]
At the candidate-generation stage, however, the material-dependent anisotropy terms required for a direct continuous minimization are not known. 
We therefore introduce a high-symmetry-direction approximation: for each selected irrep and compatible MSG, we retain basis-vector combinations whose magnetic moments lie along high-symmetry crystallographic directions of the parent paramagnetic structure. The combined constraints reduce the continuous repeated-irrep freedom to a small and finite set of physically motivated candidate magnetic configurations. The effectiveness of this approximation is confirmed by the MAGNDATA benchmark in the Supplementary Information.
\subsection*{OSSG refinement of MSG-fixed branches}

The special configurations selected from an MSG-fixed branch are further analyzed using oriented spin space groups (OSSGs). An OSSG operation is written as
\[
s=\{U\parallel R\mid \mathbf{t}\},
\]
where \(U\in O(3)\) acts in spin space and \(\{R\mid\mathbf{t}\}\) acts on the atomic sites. Its representation in magnetic-moment space is
\[
D_{\mathrm{mag}}(s)
=
P\bigl(\{R\mid\mathbf{t}\}\bigr)\otimes U,
\]
where \(P\) is the site-permutation matrix. The spin-lattice-coupled operations satisfying
\[
U=\theta\,\det(R)\,R,
\qquad
\theta=\det(U),
\]
constitute the corresponding conventional MSG subgroup. An OSSG can additionally contain spin-lattice-decoupled operations, including decoupled site-symmetry operations, and therefore imposes constraints absent at the MSG level.

For a selected MSG-fixed branch, the remaining magnetic configuration can be written as
\[
\mathbf{M}=B_{i,\mathrm{MSG}}\mathbf{c},
\]
where the columns of \(B_{i,\mathrm{MSG}}\) are the basis vectors compatible with the selected MSG branch, and \(\mathbf{c}\) contains the remaining coefficients, such as those describing the relative combination of repeated copies of the same irrep.

For each candidate OSSG compatible with this MSG-fixed branch, we then impose the invariance condition
\[
\bigl[D_{\mathrm{mag}}(s)-I\bigr]
B_{i,\mathrm{MSG}}\mathbf{c}
=
0,
\qquad
\forall\, s\in \mathrm{OSSG}.
\]
The OSSG is compatible with the branch when these equations admit a nonzero solution for \(\mathbf{c}\). Among the compatible OSSGs, we retain the symmetry-maximal ones, which impose the strongest symmetry constraints on the remaining subspace.

Because OSSG allows spin and lattice transformations to be treated independently, it contains a richer set of symmetry operations than its conventional MSG subgroup and can refine a multidimensional MSG-fixed space into lower-dimensional branches. In particular, spin-lattice-decoupled site symmetry can constrain the moment on a representative site, while the remaining OSSG operations determine the complete magnetic configuration. The OSSG analysis therefore provides a symmetry-based interpretation and refinement of the configurations selected by the high-symmetry-direction approximation. For structurally complex systems with many magnetic degrees of freedom, repeated irreps may nevertheless remain after the maximal-OSSG construction, in which case the residual freedom is further resolved using the high-symmetry crystallographic moment-direction approximation. The formal construction and the Mn$_3$GaN and Mn$_3$Sn examples are provided in Supplementary Sections I.C and I.D.

\subsection*{Propagation vectors}

The discussion above focuses primarily on \(\mathbf{q} = 0\) magnetic structures. The same symmetry-based construction can be extended to \(\mathbf{q} \neq 0\) by constructing the magnetic representation and the corresponding symmetry-adapted basis vectors for the little group of the propagation vector. In practice, we generate candidate structures for high-symmetry \(\mathbf{q}\) points of the Brillouin zone rather than relying on arbitrary real-space supercell choices. The detailed construction and benchmark for \(\mathbf{q} \neq 0\) structures are provided in the Supplementary Information.
\subsection*{First-principles calcaulations}

All total-energy calculations are performed using the projector-augmented-wave method as implemented in the Vienna Ab initio Simulation Package (VASP)\cite{vasp1,vasp2,VASPUtypeDudarev1998,GGAU}, within spin-polarized GGA+SOC+$U$ calculations. Spin-orbit coupling is included self-consistently in the energy ranking of the candidate magnetic structures. For each material, the material-specific Hubbard parameter $U_{\mathrm{lr}}$ is determined using the linear-response method \cite{PhysRevB.71.035105}. To assess the sensitivity of the predicted magnetic ground state to the on-site interaction without relying on a single empirical value, every candidate is evaluated at $U = 0$, $U_{\mathrm{lr}}/2$, and $U_{\mathrm{lr}}$, and the candidates are ranked separately for each value of $U$.Spin-group symmetries, magnetic classifications and physical properties were analysed using the FINDSPINGROUP web platform \cite{findspingroup}. The topological properties of the magnetic materials were determined by calculating the character tables of the occupied bands with Magvasp2trace and submitting them to the Bilbao Crystallographic Server\cite{xu2020high,Elcoro2021}.

For the MAGNDATA energy benchmark, compounds containing $f$-electron elements and structures with more than 60 atoms in the primitive cell are excluded. The experimentally reported magnetic structure is regarded as correctly identified when its energy lay within 30 $\mu$eV per magnetic atom of the lowest-energy candidate for at least one of the three $U$ parameterizations. The same multi-$U$ protocol is used for the high-throughput ICSD calculations; for each $U$ value, the lowest-energy candidate is assigned as the predicted ground state and configurations within the same energy window are retained as near-degenerate states. Further computational details are provided in Supplementary Section II.B.

\newpage

\bibliographystyle{aps}
\bibliography{Mstr}


\end{document}